\documentclass[utf8]{frontiersSCNS} 
\usepackage[onehalfspacing]{setspace}

\def\kms{km s$^{-1}$}

\def\keyFont{\fontsize{8}{11}\helveticabold }
\def\firstAuthorLast{Bon, E. {et~al.}} 
\def\Authors{E. Bon\,$^{1,*}$, P. Jovanovi\'c$^{1,*}$, P. Marziani\,$^{2}$, N. Bon\,$^{1}$,  \, and A. Ota\v sevi\'c}
\def\Address{$^{1}$Astronomical Observatory, Belgrade, Serbia \\
$^{2}$ Osservatorio Astronomico di Padova (INAF) Padua, Italy }
\def\corrAuthor{Edi Bon}

\def\corrEmail{ebon@aob.bg.ac.rs}

\begin{document}
\onecolumn
\firstpage{1}


\title[Exploring  relations of variability time scales and broad emission line shapes in AGN]{Exploring  possible relations between optical variability time scales and broad emission line shapes in AGN} 


\author[\firstAuthorLast ]{\Authors} 
\address{} 
\correspondance{} 

\extraAuth{}

\maketitle

\begin{abstract}

{Here we investigate the connection of broad emission line shapes and continuum light curve variability time scales of {type-1} Active  Galactic  Nuclei (AGN).}
We developed a new model to describe optical broad emission lines as an accretion disk model of a line profile with additional ring emission. We connect ring radii with orbital time scales derived from optical light curves, and using Kepler's third law, we calculate mass of central supermassive black hole (SMBH). 
The obtained results for central black hole masses are in a good agreement with {other methods. This indicates that the variability time scales of AGN may not be stochastic, but rather connected to the orbital time scales which depend on the central SMBH mass.} 

\section{}

\tiny
 \keyFont{ \section{Keywords:} galaxies:active-galaxies, quasar:supermassive black holes, quasar:emission lines, line:profiles} 
\end{abstract}

\section{Introduction}

\indent  {Type-1 Active Galactic Nuclei (AGN) are very {powerful and variable emitters}. The  continuum emission in radio-quiet AGN is   assumed to originate mainly from an accretion disk (AD) around a central supermassive black hole (SMBH) \citep{LyndenBell1969Natur,SS1973}. There is consensus that the accretion disk can be represented as  virialized (rotating) gas  in a flattened distribution following a Keplerian velocity field.}  




The optical broad emission lines (BELs) are assumed to be produced with photoionization processes \citep{Netzer2013book}, {as they respond to the variations of UV continuum}. {The accretion disk itself may be a low-ionization line emitter \citep[see, e. g.][]{CH89,CHF1989,EraHalp1994,Pop04, Bon06,Gavr07,Bon08,bonetal09,Bon09,Bon15}, if part of the continuum radiation is scattered toward it.  }The optical broad line light curves (like for e. g. H$\alpha$) are highly correlated with the optical continuum light curves \citep[e. g.,][]{Kaspi2000}.  The time lags between correlated patterns in the continuum and  H$\alpha$ flux are of the order of days, up to months,  defining the size of broad line region (BLR).  Reverberation mapping campaigns are based on this fact, and they measure time lags of correlated light curves in order to determine the sizes of reverberating region (which are after used for determination of central BH masses) in AGN \cite[see e. g. ][]{Peterson2002,DeneyK2009ApJ,Kaspi2000,Peterson1997iagn.book}.
The correlation of light curves might indicate that the signature of the main driver of the variability could be detected in the shape of the broad emission line  profiles. This could give us the velocity resolved information about processes that drive the variability. 

\indent  The shape of Balmer emission lines, particularly H$\alpha$ and H$\beta$ line profile, in many AGN 
could be described as similar to the shape of an AD line profile expected
to originate from flattened distribution of rotating emitting gas. Characteristic double peaked very broad profile, are usually seen in simulated emission line profiles \citep[see for e. g.][]{CH89,CHF1989,Cadez98,EraHalp1994,Newman97,Pop03,Pop04,Bon06,Bon08,Bon09,bonetal09,Jov10} {but} are {observed} only in one percent of AGN \citep{EraHalp1994,Strateva2003,Netzer2013book}. In typical, single peaked BEL profiles, the AD contributions could be blended by surrounding isotropic velocity component in the BLR \citep{Pop04,Bon06,Bon08,Bon09,Gavr07,Collin06,Pop03}. 
In some cases, broad H$\alpha$\  and H$\beta$ line profiles show  a red side more broadened and extended further into the red part of the spectrum, that could be associated with gravitational redshift \citep[see, e. g.,][and references therein]{Bon15}. In case the blue side of a line is more extended, the contribution is usually assumed to be from the winds or outflows \cite[see, ][]{Czerny2006,Collin06,Marziani1996,Sul2007ApJ}.

Even though  long term monitoring campaigns of AGN may still not be long enough to search for periodic variability \citep{Bon17}, there are some highly monitored cases for which  some claims of detecting significant periodicity  have been made \citep[see for example][]{Bon12,Gra2015Natur,Graham2015,Charsi16,Bon2016,Bon17,Bhatta}.
Currently, there are some ongoing campaigns of extensive monitoring programs \cite[see e. g. ][]{Gezari2007,Sergeev2007,Peterson2002,Ilic2017FrASS} that may provide valuable light curves for the future variability investigations.

Here {we explore a possibility that} ripples in the observed broad {optical} emission line profiles {may be in connection with variability time scales observed in optical continuum} light curves. {We present a case study of Arp 102B, using spectra and light curves from} \citet{Shapovalova2013A&A...559A..10S}. 
{Our hypothesis is that the variability patterns in the light curves may be induced by the orbiting {of perturbers "perturbation" (a hot spot, a spiral arm, a compact body such as a stellar mass object or even up to intermediate mass black hole) within the accretion disk \cite[see more i, for e. g.][]{,Chakrabarti199,DLin1996,Newman97,Flohic2008ApJ,Gezari2007,Jov10,McKernan2012}}. Using the AD model with additional emitting rings ({separated with gaps}), developed specially for this purpose, we match the synthetic broad emission line profiles to the observed H$\alpha$ line, in order to measure the ring radii. We connect  variability time scales, with radii, and calculate the central BH mass}.   

The paper is organized as follows: first, we present the method (section \ref{method}) of measuring the radii that could be paired with {variability time scales} from light curves. For that purpose we developed a model of an AD emission with additional enhanced thin rings (section \ref{model}). We {used data  (spectra and light curves) available in \cite{Shapovalova2013A&A...559A..10S}}.
We match the AD model to the observed 
H$\alpha$\ 
broad emission line and measure the ring radii {that we connect to optical variability time scales}. 
We analyze optical light curves, {find variability time scales  that we match with  measure ring radii, and calculate the mass of the}   SMBH  {as a test of our hypothesis}
(section \ref{results}). We discuss possible mechanisms in section \ref{discussion}.
In the last section \ref{conclusions}, we point out the main conclusions of our investigation.


\section{Method} \label{method}

Light curves of an optical continuum and broad emission line flux are highly correlated \citep{Kaspi2000}. This may indicate   
the same origin of their variability source. {Therefore, one could expect that the source of variability could leave a trace in the shapes of their broad line profile shape}.  Analysis of variation time scales may give us valuable information about why they vary the way they do, while the line profiles could provide us with the information about the kinematic parameters of the variability drive (like  the radii where the source of variation is located).  

In order to investigate the variability {time scales} of optical continuum light curves, {we use standard methods like Lomb-Scargle} \cite{Lomb76,Scargle82}, {and sine function fitting, that we use here to determine variability} time scales. Here we assume that the variability time scales corresponds to the orbital time scales within the region of AD where the optical light could be originating from \cite[see e. g.,][]{,Bon12,Bon2016}. 
{Ripples} in the broad emission line profiles   could be produced by the effects of the same phenomena that drives the variability \cite{EraHalp1994,McKernan2013}.  {If we detect them,} we would then be able to determine some dynamical properties \cite[see for e. g.][]{Newman97,Gezari2007}. In case we could identify more then one variability time scale period in the  light curves that could be linked to the radius of an emitting  ring in the broad emission line profile, then for each ring-radius pair we should expect to obtain the same mass (or at least very close value) of the central SMBH using Kepler's laws.     

In order to test these assumptions, as a first step we model synthetic line emission of an orbiting gas in the flat, disk like gas distribution, assuming that photo ionization process produces the emission line from that region, that we could approximate with the accretion disk emission model 
\citep{CH89,CHF1989,Cadez98,EraHalp1994,Jov08For,Jov10,Bon15}. 

By matching the AD model to the {H$_{\alpha}$ broad emission} line profile, we determine the inclination, inner and outer radii and additional ring radii. {The inner radius is defined by matching the far red and blue wings of the observed line profile to the red wing of AD model line shape. The red wing part is most sensitive to the and  gravitational redshift effect \cite[see, e. g.][]{Bon15}, and therefore the extension of the red side of the line wing determines how close to the BH the gas is emitting the optical Balmer lines. We note that we determine the inner radius of an AD of the H$\alpha$\ broad line from the fit of the model, and that the inner  radius in gravitational units is usually $\gtrsim $ 100, much larger than the  inner radius derived from X-ray lines  (like Fe K$_{\alpha}$), usually close to the innermost stable circular orbit, 1 to 6 gravitational radii      \citep[for AD size of the optical emission lines, see for e. g.][]{CH89,CHF1989,EraHalp1994,Pop04, Bon06,Gavr07,Bon08,bonetal09,Bon09,Bon15}.} { The part of the disk that contributes to the thermal continuum emission is the one of highest temperature, also close to the the innermost stable circular orbit \citep{LyndenBell1969Natur,SS1973}. This region is expected to be much closer to the black hole than the AD region suitable for    the optical emission line emission. {From the fit of the model to the observed broad line we determine inner and outer radius of the AD and each ring  as well as the inclination and emissivity law that is common parameter for the complete model}.} 

Simulated profiles of AD emission usually have characteristic two peaks in the core of the line,   are broadened due to {virial motion} and made asymmetric by relativistic effects. The two peaks are usually blended by the isotropic emission component {in the core of the line profile, originated further} away from the AD, which is {observed} in majority of AGN spectra \citep{Bon06,Bon08,Bon09,Pop03,Pop04,Pop08}. Only in a very small number (less then 1\%) of objects the two peaks are clearly recognized \citep[see e. g. ][]{Strateva2003}). 

Small bump-like features can be found  on the BEL profiles. They cannot be immediately modeled by smooth AD profiles.

\subsection{Model of AD with additional ring emission or AD with ring gaps} \label{model}

The AD model is an idealization of emission with assumption of an homogeneous AD. This may not be the case, and therefore  the AD profile may  not sufficient to describe all features in observed profiles like e. g., small bumps on the wings which are often present. 

Assuming that the time scale of perturbed disk (cooling time, shock wave progression, or anything that produced additional emission from that ring) is significantly longer then the orbital time scale, then we could approximate that the {time scale of} variations {measured} in optical light curves correspond to the orbiting of some features {within} the AD at radii that {could be} associated to  {narrow} rings that we {located in AD} by matching the {observed} emission lines with synthetic modeled profiles. Their radii are measured in units of gravitational radii Rg, since from the AD model we cannot obtain the information about the central mass. 
By connecting each {variability time scale} to {some radii} in the AD, with an assumption that the shorter {variability time scale} corresponds to a closer ring, while the outer radii correspond to a longer {variability time scale}, we could be able to calculate the mass of the central SMBH  \footnote{This is similar as in the case of planetary system where all orbiting objects radii and periods correspond to the same mass of the central body} using the Kepler's third law for a circular orbit:

\begin{equation} 
P=2 \pi \sqrt{r^{3}\over{GM}}={{2 \pi G M \xi^{3/2}} \over{c^{3}}}, \label{eq}
\end{equation} 

where  $r=\xi (r_{g})$ is the ring radius in gravitational radii and $P$ is the circular orbital period of the orbiting region at such radii \citep[as prposed in ][]{Newman97,Gezari2007}. {We note here that regardless of the formula, our method is different then the method of \citet{Newman97}. These authors  used monitoring spectra to determine the radius of a hot spot, which they then connected to the orbiting period. Here we use single epoch spectra and variability time scales measured from photometric data.} 

 We use different AD model than in previous papers \citep{CH89,CHF1989,Gezari2007,Newman97,1996ApJ...456L..25A}, even though we obtain similar values of parameters for the inclinations and the inner  and outer radius. Here we use the relativistic ray tracing AD model\footnote{We tested several different AD models \cite{Cadez98,Jov08For,Fanton97,Jov12,Bon15} for a line fit, and find that obtained inclinations were practically the same regardless of the model used and also if  
we adopt more complex model, assuming the additional rings with the same AD parameters (inclination and emissivity law).} We propose that the origin of the variability patterns, could be traced to the ripples in the {shape of} broad line profiles. Making a connection of {variability time scales} and radii, Eq. \ref{eq} can {be used to determine} dynamical properties of AGN. 


{We construct our} model assuming the emission of an AD 
and each ring 
contributions to the line profiles with the same {inclination and emissivity law} as in the parent AD, which is preserved in ring models.
	The code  includes 
	both special relativistic and general relativistic effects on radiation from the 
	accretion disk around SMBH \citep[see e.g.][]{Jov12}. This AD model is based on ray-tracing 
	method in the Kerr metric \citep{Fanton97,Cadez98}, for different values of inner and outer radii and inclinations of rings in AD.  
The emissivity index was {kept to be close as possible to the value} q=-2, assuming the emissivity law to be $\sim r^{q}$, as expected for the case of photo-ionization mechanism. The model is then constructed using a previous match of the AD profile to the emission line, as a starting point. The scaled contributions of the ring profiles are added to the AD profile until bumpy features  in observed spectra are described with the synthetic spectrum. Beside the fact that the shape of the line is fitted more realistic then with a  simple AD model, we are also obtaining a valuable information about the radii in the disk plane where the emission is emitted from.

{
	We considered 
	our model as analogy to AD
	in which the emission is not continuous from $R_{min}$ to $R_{max}$, but instead is restricted to emission emission annuli. 
	The physical justification of disk gaps can reside on the removal of the gas disk in systems in which the {the emission is not contributing to the emission line shape due to obscuration by dust in the outscirt of BLR \cite{Bon2016} or by voids produced due to accreting secondary object within the AD plane \cite{DLin1996,DLin1997}, or in case where} gas was originally counterrotating with respect to the black hole spin (Vorobyov et al. 2016), or by an accreting object in the disk plane (Artymovicz and Lin 1993). In this case, the disk profile is modeled by the addition of several emission annuli (assuming the same AD model and one inclination value).  The analysis of the line profile is otherwise analogous to the excess ring emission model. }

\section{Results} \label{results}

In order to test this our model interpretation of the BEL profiles, we 
{performed a case study of Arp 102B
because it is well known for its broad emission lines}
with double peaked shape that was already proposed to correspond to the AD emission profile \citep{CH89,CHF1989,EraHalp1994,Gezari2007,Jov10,Newman97,PopShap2014A&A...572A..66P,Sergeev2000A&A...356...41S,1996ApJ...456L..25A,1990ApJ...355L..15S}.   

Using the line profile matching to the model, we measure the ring radii that we connect to the variability {time scales}.

Assuming that the orbital time scale is the only match to time scale of variability patterns seen in these light curves, we combine measured radii  $\xi$ and {variability scale} periods, and derive mass assuming circular Keplerian orbit of this {variable} region positioned within the AD.\footnote{It is important to stress that emission lines do respond to continuum changes on a timescale that is $\tau \ll P$\ (months vs. 10 years), but this should not invalidate our argument right because the reverberation response $\tau$ is much smaller than $P$, which is the orbital period at radius $\xi$.}  

We measured $\xi$ radii {with an AD model} in units of gravitational radii (R$_g$) from the {broad emission} line profiles. If two or more significant periods were detected, {assuming a low cut taken to be at} 400 days period (in order to avoid effects of Earth's orbital period of 1 year), {then we try to pair the to the parts of an AD which could be perturbed, like inner and outer radius of a ring or a main part of AD}, {where the interaction with different state medium could be expected}.\footnote{We note that the expected orbital periods in optical part of the AD for typical AGN  of M $\sim 10^8$ M$_{\odot}$ should be about 1 year \citep[see for e. g.][]{Gaskell2008RMxAC}, but this should not be a problem since also longer periods are detected.}  We test the assumption by calculating the mass of the central BH following Eq. \ref{eq}, with expectation that the obtained results for masses, for each pair of {variability time scale} and {the ring} radius, should be equal, or at least to be of similar values.

\begin{figure}[h!]
	\begin{center}
		\includegraphics[angle=270,width=12cm ]{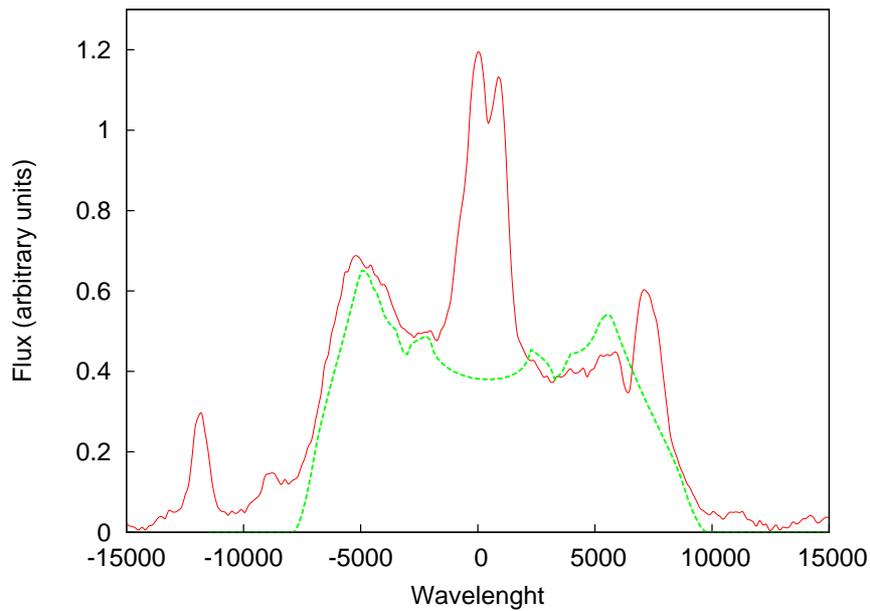}
		
	\end{center}
	\caption{An example of AD + rings model matching the observed  broad H${\alpha}$ emission line profiles. Here is presented the case of Arp 102B}\label{fig:examples}
\end{figure}

\begin{figure}
	\begin{center}
		\includegraphics[angle=270,width=12cm ]{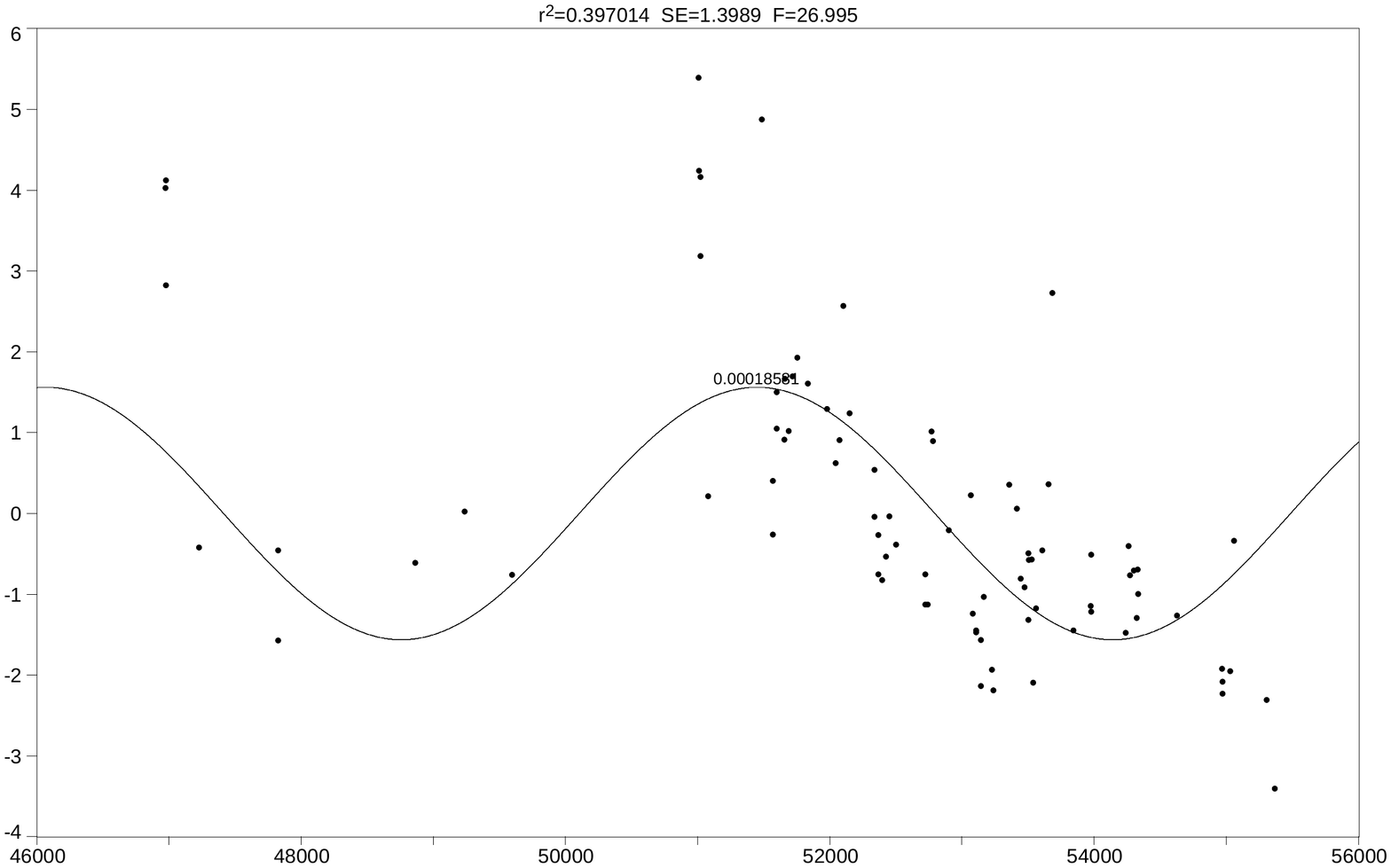}
	\end{center}
	\caption{Long term variability analysis of Arp 102B light curve after trend removal correction {fitted with the sine function with the period} of about 14 years.}\label{fig:time}
\end{figure}

The example {of the model matched to the observed spectrum of the object Arp 102B is presented in Fig. \ref{fig:examples}. The inclination is 32 degrees, and the AD is truncated in the following rings:  370-1050 Rg,  1430-1540 Rg and 3500-3800 Rg, with the emissivity parameter q=-2.1 {over the whole model}. We note that the result is very similar to other fits of this object \cite{CH89, CHF1989, Newman97}, who obtained the same inclination while the complete disk was assumed to correspond to main part of AD in our model, with Rinn = 350 Rg, and Rout = 1000 Rg\footnote{We note that at about the same radius was obtained as outer radius of AD in previous disk models \cite{CH89,CHF1989,Newman97}}.} {The measurement of variability time scales of light curves is shown in the Fig. \ref{fig:time}. We find the longer time scale to be about 14 years period (intrinsically 5250 days, after correction for the systematic redshift), while the shorter time scale is taken from earlier variability analysis \cite{Newman97, Sergeev2000A&A...356...41S, Shapovalova2013A&A...559A..10S}, where the variability time scale was found to be of about 700 days (in the interval 650-750 days, here taken as 680 after scaling for the redshift).}

We assume that the variability is produced {sharp edges of gaps and emitting rings}.
By matching the inner radius of AD to the corresponding variability time scale (370 Rg with 680 days), and 1500 Rg with 5255 days variability time scale), we obtain the mass of about 2.7 $\cdot 10^{8}$ M$\odot$ (as an average of these two values: 2.68 $\cdot 10^{8}$ and 2.71  $\cdot 10^{8}$ M$\odot$).
This value is very close to the value of with 2.1 $\cdot 10^{8}$ M$\odot$ in \citet[][obtained from the orbiting hotspot assumption]{Newman97}
and not so different from
 3.5 $\cdot 10^{8}$ M$\odot$ suggested in \citet[][obtained with the model of thousands randum orbiting clouds, mached to the radial velocity maps of broad line variability]{Sergeev2000A&A...356...41S},  and about 2 times higher then the mass value of 1.1  $\cdot 10^{8}$ M$\odot$ as suggested in \citet{Shapovalova2013A&A...559A..10S}. 

{The virial product $r \delta{v_{r}}^{2}/G$ can then be obtained from the   20 day lag of line flux to continuum \citet{Shapovalova2013A&A...559A..10S}}, and { following the formula form \cite{Collin06} for radial velocity dispersion in case of rectangular shape\footnote{line $FWHM / \sigma = 2\times \sqrt{3} = 3.46$} line profile $\delta{v_{r}}$ = 14500 / 3.46 $\approx$ 4190 \kms. The virial factor $f$ can then be estimated by computing $r \delta{v_{r}}^{2}/G/M_{BH,P}$ = $1/f$, where $M_{BH,P}$\ is the black hole mass computed from the $P$ following Eq. \ref{eq}. Using these values we derive $1/f$ = $ 7.36 \cdot 10^{7} /(2.7 \cdot 10^{8}) \approx 0.287.$.      The $1/f$ {should be} equal to ratio of virial product and the mass\footnote{$f = ((\sin(i))^{2} + (H/R)^2)^{-1}$ as suggested in \citet{Netzer2013book}, were H/R represents the contribution of isotropic velocity component as a ratio between the width of a disk and the radius of AD. Assuming H/R $\approx$ 0 the formula transforms into $f =(sin(i))^{-2}$}.  obtained from the period by definition, and $f$ depends on inclination as $f \approx (sin(i))^{-2}$, resulting the inclination of 32{$^\circ$},} 
which is 
{consistent with the value} from the AD model fitted to the H$\alpha$ broad emission line (with our model and earlier fits as well). Therefore, not only that the model gives the consistent mass with the three different period-radii pairs, but also the retrieved inclination from the virial product gives a consistent value of inclination as the AD model fit to the H$\alpha$ broad emission line.

{Admittedly, our model follow from a very strong assumption: the use of variability time scale $P$ (derived from the continuum variation) for estimating the black hole mass and the use of $\xi$ derived from the profile ring should be consistent. The assumption may not be correct in principle, since the $P$\ obtained from the continuum may not refer to the same $\xi$\ of the profile rings. It requires  in-situ emission at the $\xi$\ deduced from the ring. }

\section{Possible interpretations} \label{discussion}

Here we do not consider any details about what produces the hot ring regions or the dips in the AD, and we are fully aware that the measured periods are significant above the white noise levels, but may not appear significant compared to the red noise AR curves \citep{Vaughan06,Vaughan16}. We were mainly interested in measuring time scales of orbital periods assuming that the variability patterns \cite{Marz2017pattern,Marz17IAUS} in the light curves could be induced by the orbital time scales.
{Some} interpretations of {possible} periodicities are discussed in many works \citep{Bon2016,Charsi16,Chakrabarti199,EraHalp1994,Graham2015,Gezari2007,Gra2015Natur,Liu16,Li2016,Newman97,Bon17,Pop2012NewAR}. Assuming circular orbits in the disk as we did here, we suggest that possible source of {optical variability} should be located in the AD, amplifying emission at that radius. We are aware that at such radii, the standard models of thermal emission of AD \citep{SS1973} shows that the temperature of the disk is relatively low, under 1000 K {\cite[according to the standard disk model,][]{SS1973}}, which is not sufficient enough for the photo ionization mechanism to produce optical broad emission lines, or to significantly contribute to the optical continuum flux, without additional emission mechanism, like shocks \citep{Chakrabarti199,EraHalp1994,Gezari2007}, hot spot \cite{Jov10,Newman97,Flohic2008ApJ}, secondary orbiting object on a circular orbit around the central SMBH with additional accretion mechanism that is sufficient to produce significant contribution to the continuum and the line emission \cite[see e. g.][where they show a fast forming of intermediate mass BH's in AD with circular orbits arround the cetral SMBH]{McKernan2014MNRAS,DLin1996,DLin1997}). It is expected that in such cases the voids or gaps would be formed \citep[see, for e. g., ][]{McKernan2012,McKernan2013,McKernan2014MNRAS} in the AD (similar to the planet formation in the stellar disks), with piling up of matter at the outer border of the gap ring \cite[see,][and the references therein]{McKernan2013}, that may {represent} the region that could be associated with the ring emission that we modeled here.

\section{Conclusions} \label{conclusions}

We simulated an AD emission profiles with {additional} ring regions and compared them with the observed profiles of the broad H$\alpha$ emission lines.  We {measured variability time scales} from available optical light curves. We pair each ring profile with the {variability time scale}. Using Kepler's third law, we then calculate central SMBH masses. 

Our results show that:

1) The model of an AD with {additional emitting rings}  {or by a sequence of emission annuli separated by gaps} could well describe the observed H$\alpha$ emission line profiles in the case of Arp 102B.

2) Masses calculated from each pair of {variability time scale} and {narrow} ring {with enhanced emission}, result with very similar values 
{providing support for the model and initial hypothesis, indicating that the features fitted in the line profiles are probably emitted in an AD (virial velocity field).}

{3) We also test our model prediction from the fact that the virial factor depends on the inclination. Since, from the ratio of VP and the mass obtained with our model using Keplerian velocity, we obtain practically identical value of inclination as from the AD fit to the line profile. Therefore this gives another justification of our model.}

{This result may indicate that the variability time scales of AGN may be connected to the orbiting time scales which depend on the central SMBH mass.}

{In near future}, we plan to extend the sample (Bon et al. 2018, in preparation), selecting more AGN with long term monitoring data.

\section*{Conflict of Interest Statement}
The authors declare that the research was conducted in the absence of any commercial or financial relationships that could be construed as a potential conflict of interest.
 
\section*{Author Contributions}
EB is responsible for developing the idea and writing the text of the manuscript. PJ developed the code for the model. EB, NB, PJ, PM and OA worked on analyses and discussions

\section*{Funding}
This research is part of projects 
176003 “Gravitation and the large scale structure of the
Universe” and 176001 “Astrophysical spectroscopy of extragalactic
objects” supported by the Ministry of Education and
Science of the Republic of Serbia. 

\section*{Acknowledgments}
This research is part of projects 
176003 “Gravitation and the large scale structure of the
Universe” and 176001 “Astrophysical spectroscopy of extragalactic
objects” supported by the Ministry of Education and
Science of the Republic of Serbia. 

\bibliographystyle{frontiersinHLTH&FPHY} 
\bibliography{EBon_ref}


\section*{Figure captions}


\begin{figure}[h!]
	\begin{center}
		\includegraphics[angle=270, width=16cm ]{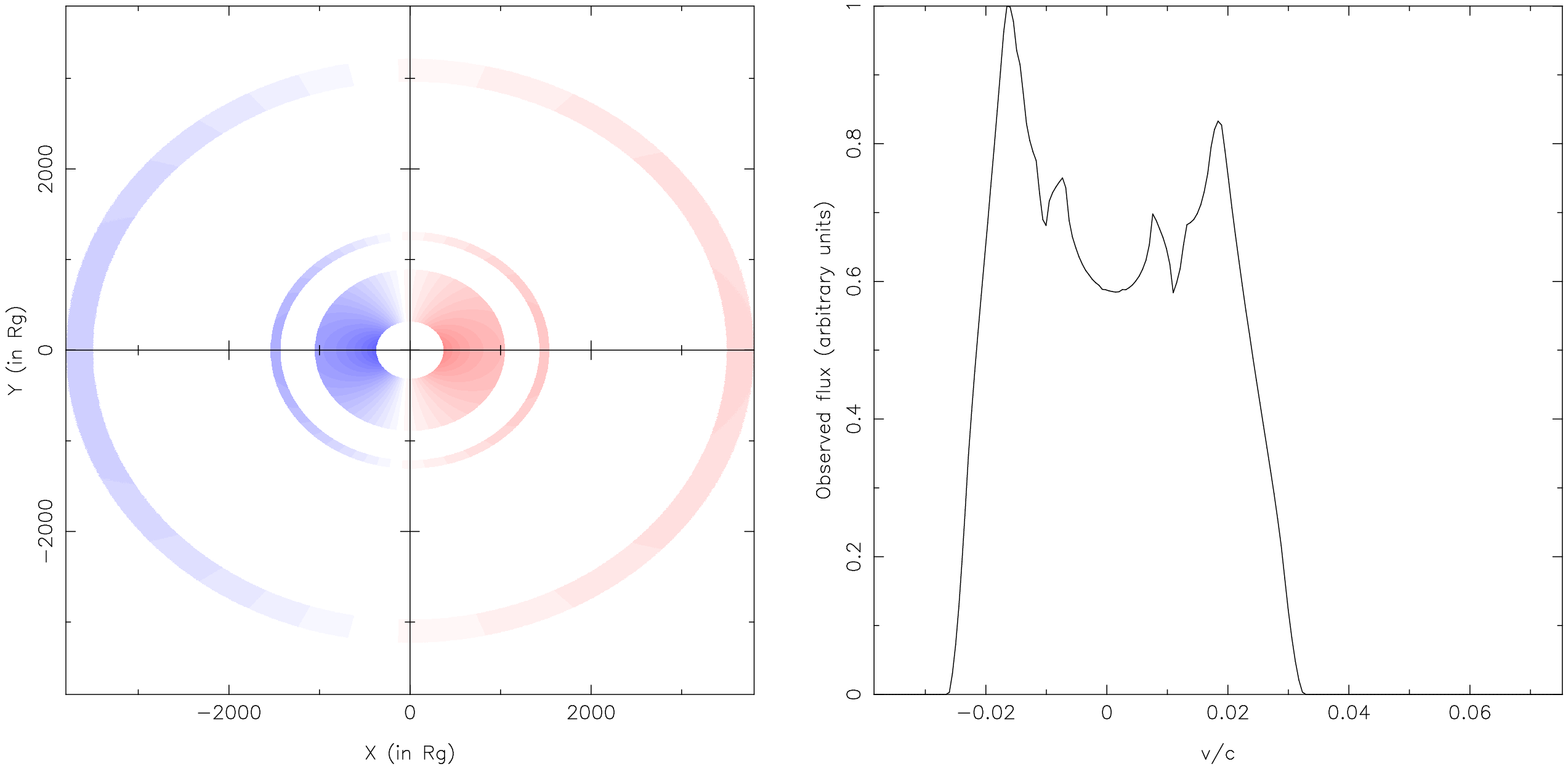}
	\end{center}
	\caption{An examples of AD + 2 outer rings (left panel) with corresponding shape of broad line profile, matched to H$\alpha$ line of Arp 102B in \ref{fig:examples}. }\label{fig:examples2}
\end{figure}

\end{document}